\documentstyle{mn}

\newcommand{\etal}{{et al.}}

\newcommand{\kev}{\,\mbox{${\,\rm keV}$}}

\newcommand{\msunyr}{\mbox{\,{\rm M}$_\odot$\ {\rm yr}$^{-1}$\,}}
\newcommand{\ergs}{\mbox{${\,\rm erg~s}^{-1}$\,}}

\newcommand{\peryr}{\mbox{${\,\rm yr}^{-1}$\,}}
\newcommand{\cmcub}{\mbox{${\,\rm cm}^{-3}$\,}}

\newcommand{\kt}{\mbox{$kT$}}

\newcommand{\rosat}{\mbox{\sl ROSAT}}
\newcommand{\einstein}{\mbox{\sl EINSTEIN}}
\newcommand{\chandra}{\mbox{\sl Chandra}}

\newcommand{\hii}{\mbox{{H{\small II}}}}

\title[Radio Observations of Super Star Clusters in Dwarf
Starburst Galaxies] {Radio Observations of Super Star
Clusters in Dwarf Starburst Galaxies}

\author[I.\,R. Stevens, D.\,A. Forbes, R.\,P. Norris]{Ian R. Stevens$^{1}$, 
Duncan A. Forbes$^{2}$,  Ray P. Norris$^{3}$\\ 
$^{1}$ School of Physics and Astronomy, University of Birmingham, 
Edgbaston, Birmingham, B15 2TT, UK\\
$^{2}$ Centre for Astrophysics and Supercomputing, Swinburne University,
Hawthorn VIC 3122, Australia\\
$^{3}$ Australia Telescope National Facility, CSIRO, PO Box 76,
Epping, NSW 2121, Australia}

\date{Accepted ..............................; 
Received ..............................; 
in original form ..............................}

\begin{document}

\maketitle

\begin{abstract} 

We present new radio continuum observations of two dwarf starburst
galaxies, NGC\,3125 and NGC\,5408, with observations at 4.80GHz (6cm)
and 8.64GHz (3cm), taken with the Australia Telescope Compact Array
(ATCA).  Both galaxies show a complex radio morphology with several
emission regions, mostly coincident with massive young star
clusters. The radio spectral indices of these regions are negative (with
$\alpha\sim -0.5 - -0.7$), indicating that the radio emission is
dominated by synchrotron emission associated with supernova activity
from the starburst. One emission region in NGC\,5408 has a flatter index
($\alpha \sim -0.1$) indicative of optically thin free-free emission,
which could indicate it is a younger cluster.
Consequently, in these galaxies we do not see regions with the
characteristic positive spectral index indicative of optically obscured
star-formation regions, as seen in other dwarf starbursts such as
Hen~2-10.

\end{abstract}


\section{Introduction}
\label{sec1}

Dwarf starburst galaxies are the more numerous and fainter counterparts
of traditional starburst galaxies.  Dwarf starbursts will have a
particularly important role in the energizing and polluting of the
intergalactic medium through galactic winds, as, because of their low
escape velocities, even a moderate starburst event in a dwarf galaxy can
result in the expulsion of a large fraction of the nuclear processed
material in a superwind or outflow (Dekel \& Silk 1986; Mac Low \&
Ferrara 1999; Strickland \& Stevens 2000). Understanding the
complexities of starburst processes in dwarf galaxies is therefore an
important step in understanding the feedback processes between
star-formation and galactic evolution.

Radio continuum observations of starburst galaxies are particularly
useful in studying the star-formation processes, suffering little of the
extinction that complicates matters in the optical and UV, which may
appear to be more suitable wavebands for studying young massive star
populations (see Hopkins \etal\ 2001 for a discussion of the merits of
different wavelength regimes for finding star formation rates). Radio
observations can also be used to study populations of supernovae
remnants (SNRs) in starburst galaxies (e.g. Kronberg \etal\ 2000, Greve
\etal\ 2002), and recent radio observations have revealed objects
inferred to be optically obscured super star clusters (see below).

Earlier radio observations of dwarf starbursts, at low spatial
resolution, have shown them to have a higher radio to optical luminosity
ratio than normal spiral galaxies and spectral indices significantly
flatter than a sample of spiral galaxies (Klein, Wielebinski \& Thuan
1984). This was attributed to lower magnetic fields in the dwarf
galaxies effectively suppressing non-thermal emission.  However, more
recent and higher resolution observations are showing a more complex and
interesting picture. VLA observations of the dwarf starburst Henize~2-10
by Kobulnicky \& Johnson (1999, see also Kobulnicky \& Johnson 2000)
found several compact radio sources in the central regions. These
sources have positive spectral indices (i.e. $\alpha>0$, with the
radio flux defined as $S_\nu\propto \nu^\alpha$), indicative of
optically thick bremsstrahlung emission. These radio knots are believed
to be unusually dense \hii\ regions, with sizes of $\sim 3-8$pc and
densities $\sim 1500-5000\cmcub$, associated with optically observed
super star clusters.

Further, VLA observations of NGC\,5253 (another dwarf starburst) showed
unusually high levels of free-free emission, most of it from a single
source (Turner, Ho \& Beck 1998).  Several regions with a positive
spectral index were also detected, which were inferred to be very young
regions of star-forming activity. These authors also found regions of
non-thermal emission, probably associated with SNRs. Further
observations showed what was inferred to be a dense compact \hii\
region, with a size of only 1--2pc, but containing several thousand
O-stars, and which was termed a \lq\lq radio supernebula\rq\rq\ (Turner,
Beck \& Ho 2000).

Beck, Turner \& Kovo (2000) have presented VLA observations of nine
Wolf-Rayet galaxies (a particular class of starburst galaxies,
containing very young and massive Wolf-Rayet stars, Conti 1991),
including seven dwarf starbursts. In the majority of these galaxies the
global spectral indices were much flatter than in spiral galaxies (the
only exception being NGC\,3049 which is a large spiral galaxy as well
as a Wolf-Rayet galaxy, which has a spectral index of $-0.7$). This
flatter index indicates a higher fraction of free-free emission in these
galaxies, and in fact many of the galaxies had regions with positive
spectral index, indicative of self-absorbed free-free emission and hence
very recent and compact star-formation.  These objects are probably
related to the optical super star clusters (SSCs) observed with {\sl
HST} in many starbursts and galaxy mergers (see Whitmore 2002 for a
review), but which are viewed at a younger age, and which are still
enclosed within the parental cloud, obscuring them at optical
wavelengths.

Other examples of optically obscured SSCs seen at radio wavelengths are
found in NGC\,2146 (a spiral starburst galaxy), where a number of radio
sources with positive spectral indices are seen, and that have been
interpreted as being due to massive ultra-compact \hii\ regions,
indicative of extremely young star-formation (Tarchi \etal\
2000). Kobulnicky \& Johnson (2001) and Johnson \etal\ (2001) summarise
the status of radio observations of these young starburst regions, and
coins the phrase \lq\lq ultra-dense\rq\rq\ \hii\ regions (or UD\hii\
regions). It is worth noting that {\sl MERLIN} observations of the
nearby (2.2Mpc) dwarf starburst NGC\,1569 found no emission associated
with the optically visible SSCs in this galaxy, but did find a number of
radio supernovae or supernovae remnants (Greve \etal\ 2002).

In this paper we report on radio observations, taken with the Australia
Telescope Compact Array (ATCA), of two dwarf starburst galaxies,
NGC\,3125 and NGC\,5408, which are also Wolf-Rayet galaxies. Because
Wolf-Rayet galaxies contain very young massive stars and are thus sites
of very recent star-formation, they are the obvious place to look for
optically obscured star clusters. Radio maps with good spatial
resolutions are therefore very important in understanding the global
properties of these objects.

The paper is organized as follows: in Section~2 the characteristics of
the target galaxies are discussed, in Section~3 the radio observations
and analysis are described and the results are discussed in Section~4.

\begin{table}
\caption{Parameters for NGC\,3125 and NGC\,5408. }
\begin{tabular}{lcc}\hline
                     & NGC\,3125 & NGC\,5408 \\ \hline
Classification       & BCDG      &  IB(s)m    \\
RA  (J2000)          & $10^h06^m33^s$ & $14^h03^m19^s$ \\
Dec (J2000)          & $-29^\circ56' 08''$ & $-41^\circ 23' 18''$ \\
$B_T$                & 13.50 & 12.2\\
$(B-V)_T$            & 0.50 & 0.56\\
Distance $D$ (Mpc)   & 13.8 & 8.0 \\
Optical extent  & $1.1'\times 0.7'$ & $1.6'\times 0.8'$\\ \hline
\end{tabular}
\label{tab1}
\end{table}

\section{The Target Galaxies}
\label{sec2}

\subsection{NGC\,3125 (Tol 3, Tol 1004-296)}

NGC\,3125 is a dwarf starburst and has been classified as a Wolf-Rayet
galaxy (Conti 1991). We assume a distance of 13.8\,Mpc (Marlowe \etal\
1995). In general terms NGC\,3125 is a metal-poor, irregular amorphous
dwarf galaxy, undergoing strong star-formation activity. The stellar
population seems to be the result of periodic bursts of strong star
formation followed by periods of quiescence (Kunth, Maurogordato \&
Vigroux 1988).  A spectral analysis by Vacca \& Conti (1992) suggested
that the emission line spectra required the presence of $\sim 500$ WN
type Wolf-Rayet stars and $\sim 2500$ OB stars. Population synthesis by
Raimann \etal\ (2000) suggests the presence of stars with ages ranging
from $2-3$Myr up to $>500$Myr. Their is no evidence of a recent merger
for this galaxy.

Marlowe \etal\ (1995) and Marlowe, Meurer \& Heckman (1999) observed
NGC\,3125 at optical wavelengths, as part of a study of several dwarf
starbursts. H$\alpha$ observations of NGC\,3125 show bright emission
from the central region, with a spatial extent of $25''$ ($\sim
1.7$~kpc) with three emission peaks in this region. In addition,
filamentary H$\alpha$ emission extending out to a radius of $40''$ was
seen, as well as evidence of an outflow. NGC\,3125 has also been
observed with the Faint Object Camera on the pre-refurbishment {\sl
HST}, showing several SSCs in the central regions of the galaxy.

NGC\,3125 has been observed with the \einstein\ and \rosat\ X-ray
satellites (Stevens \& Strickland 1998b). Although there is limited
spectral information, on the basis of its likely X-ray properties
(i.e. assuming that its X-ray spectrum is similar to that of other dwarf
starbursts) it is the one of most X-ray over-luminous dwarf starbursts
as compared to its B-band luminosity (Stevens \& Strickland 1998a;
1998b), with an X-ray luminosity of $L_X\sim 2\times 10^{39}\ergs$, in
the $0.1-2.5\kev$ waveband.

\subsection{NGC\,5408}

NGC\,5408 is a star-bursting dwarf irregular galaxy, and has three main
nuclear \hii\ regions (Bohuski \etal\ 1972). Colour-wise, like NGC\,3125,
it is extremely blue, (indicative of very vigorous star-formation). We
assume a distance of 8~Mpc (Fabian \& Ward 1993). Again, there is no
indication of a recent merger that might have triggered the starburst.

NGC\,5408 has been observed on several occasions at X-ray energies, and
is very X-ray luminous for a dwarf galaxy (Fabian \& Ward 1993; Stevens
\& Strickland 1998b). The X-ray morphology of NGC\,5408, as seen by the
{\sl ROSAT PSPC} shows that the X-ray emission is broadly centred on the
main nuclear \hii\ regions.  The best-fit model for the X-ray spectra has
$\kt\sim 0.5\kev$ with little local absorption. The X-ray luminosity
(corrected for absorption) is $L_X\sim 3\times 10^{40}\ergs$
($0.1-2.5\kev$).

NGC\,5408 has also been observed at higher spatial resolution with the
\rosat\ HRI (Fourniol, Pakull \& Motch 1996), showing that the X-ray
emission is dominated by a single point-like source, which is not
obviously co-located with any of the SSCs in NGC\,5408.  The origin of
the X-ray emission is rather unclear, possibly a single X-ray luminous
SN, thermal emission associated with \hii\ regions, an ultraluminous
X-ray binary or even a background quasar.

\begin{table*}
\caption{A Summary of the ATCA Radio Observations of NGC\,3125 and NGC\,5408}
\begin{tabular}{lccccc} \hline
              & \multicolumn{3}{c}{NGC\,3125} &\ \ \ \  &NGC\,5408 \\
              & \multicolumn{3}{c}{\hrulefill} & &\hrulefill \\
Date          & 25 March 2000 & 31 March 2000 & 25 October 2000 & & 31 March 2000 \\
Configuration & 6D     & 6D     & 6C     & &   6D\\
RA  (J2000) & \multicolumn{3}{c}{$10^h 06^m 33^s$}& & $14^h 03^m 19^s$ \\
Dec (J2000) & \multicolumn{3}{c}{$-29^\circ 56'08''$}& & $-41^\circ 23' 18''$\\
Frequencies$^{1}$ &\multicolumn{3}{c}{4.80GHz, 8.64GHz} && 4.80GHz, 8.64GHz\\
Total Bandwidth$^{1}$  & \multicolumn{3}{c}{128MHz} && 128MHz\\
Beam Size & \multicolumn{3}{c}{$3.34''\times 1.50''$ (8.64GHz)} & &
$1.60''\times 0.97''$ (8.64GHz)\\
 & \multicolumn{3}{c}{$6.06''\times 2.66''$ (4.80GHz) }&&
$2.90''\times 1.76''$ (4.80GHz)\\
Flux Calibrator  & \multicolumn{3}{c}{1934--638} && 1934--638\\
Phase Calibrator & \multicolumn{3}{c}{1016--311} && 1349--439\\  \hline
\end{tabular}
\begin{flushleft}
Notes on Table:\\
$^{1}$: The same for all observations\\
\end{flushleft}
\label{tab2}
\end{table*}

\section{The Radio Continuum Observations}
\label{sec3}

NGC\,3125 and NGC\,5408 were observed with the Australia Telescope
Compact Array (ATCA), located at Narrabri, New South Wales,
Australia. The observations were performed on 25 March 2000 and 31 March
2000, with the array in the 6km (6D) configuration. The observations
were performed simultaneously at 4.80GHz (6cm) and 8.64GHz
(3cm). Observations of the target galaxy and a phase calibrator were
alternated through the 12hr observing run, giving a total of about 10hr
spent on source.  The secondary flux calibrators used in these
observations were 1016--311 for NGC\,3125 and 1349--439 for NGC\,5408,
along with the standard ATCA primary calibrator, 1934--638. The flux
densities for the various calibrators were as follows; for the primary
calibrator 1934--638, 2.842Jy (8.64GHz) and 5.829Jy (4.80GHz). For the
secondary calibrator the flux densities were; for 1016--311, 0.592Jy
(8.64GHz) and 0.583Jy (4.80GHz), for 1349--439, 0.394Jy (8.64GHz) and
0.452Jy (4.80GHz).

During the first run looking at NGC\,3125 on 25 March 2000 problems with
one of the antennas was encountered, restricting the $(u,v)$ coverage
somewhat and some further scans on NGC\,3125 were performed during the
second run, which concentrated on NGC\,5408. In addition, some further
scans were made of NGC\,3125, using the 6km (6C) array, during a
subsequent run (25 October 2000).

The ATCA data were processed using the {\small MIRIAD} package (Sault,
Teuben \& Wright 1995). The data were calibrated and cleaned using
standard techniques, with the inversion done using uniform weighting. To
estimate the spectral indices of the emission regions we have convolved
the higher resolution (8.64GHz) map to match that of the lower
resolution (4.8GHz) image. Flux errors were estimated as in Venturi
\etal\ (2001), and errors on the spectral index were estimated using
standard propagation of error expressions.

\section{Results and Discussion}
\label{sec4}

\subsection{NGC\,3125}
\label{sec4p1}

\begin{figure*}
\vspace{9cm}
\includegraphics{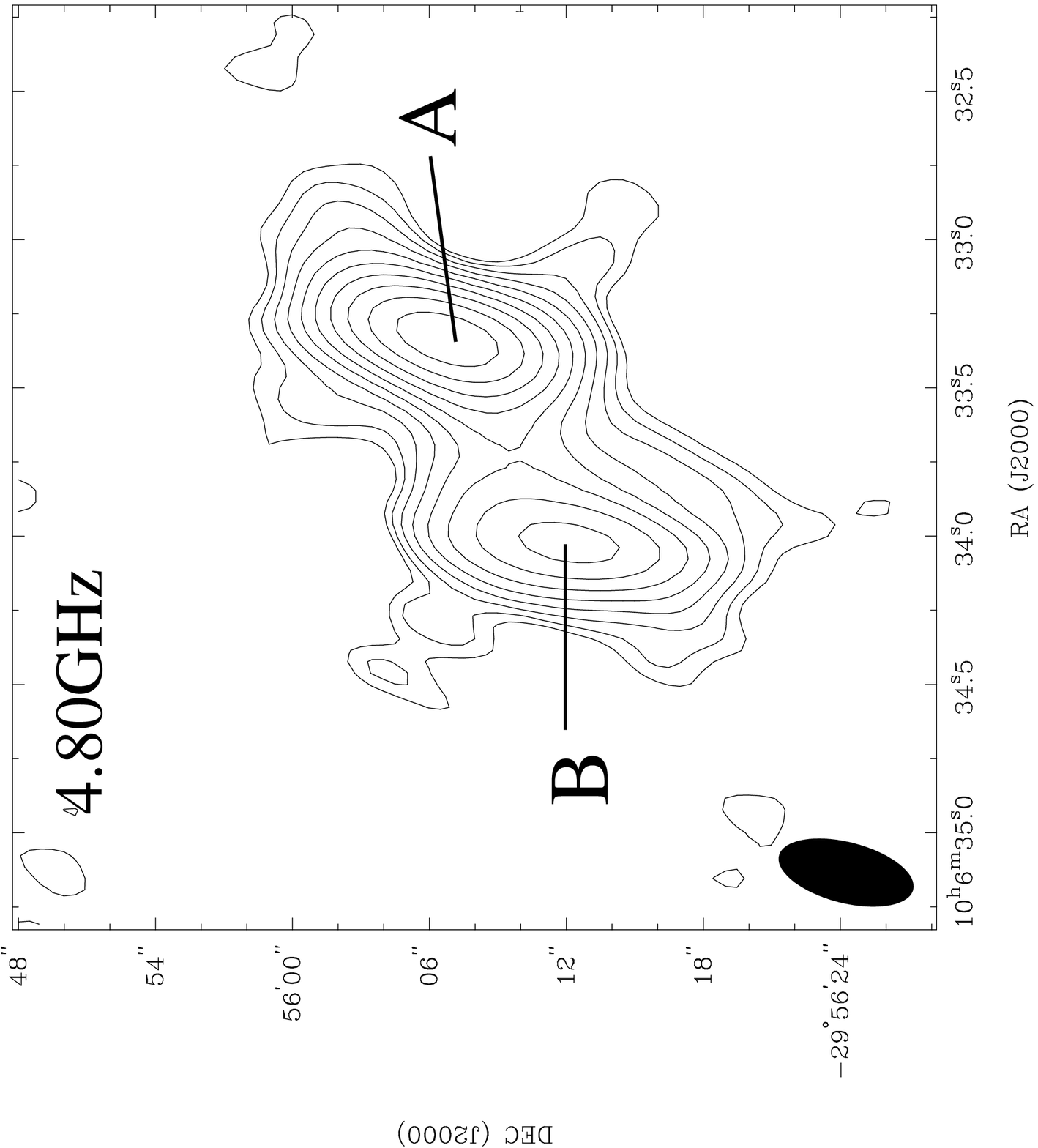}
\includegraphics{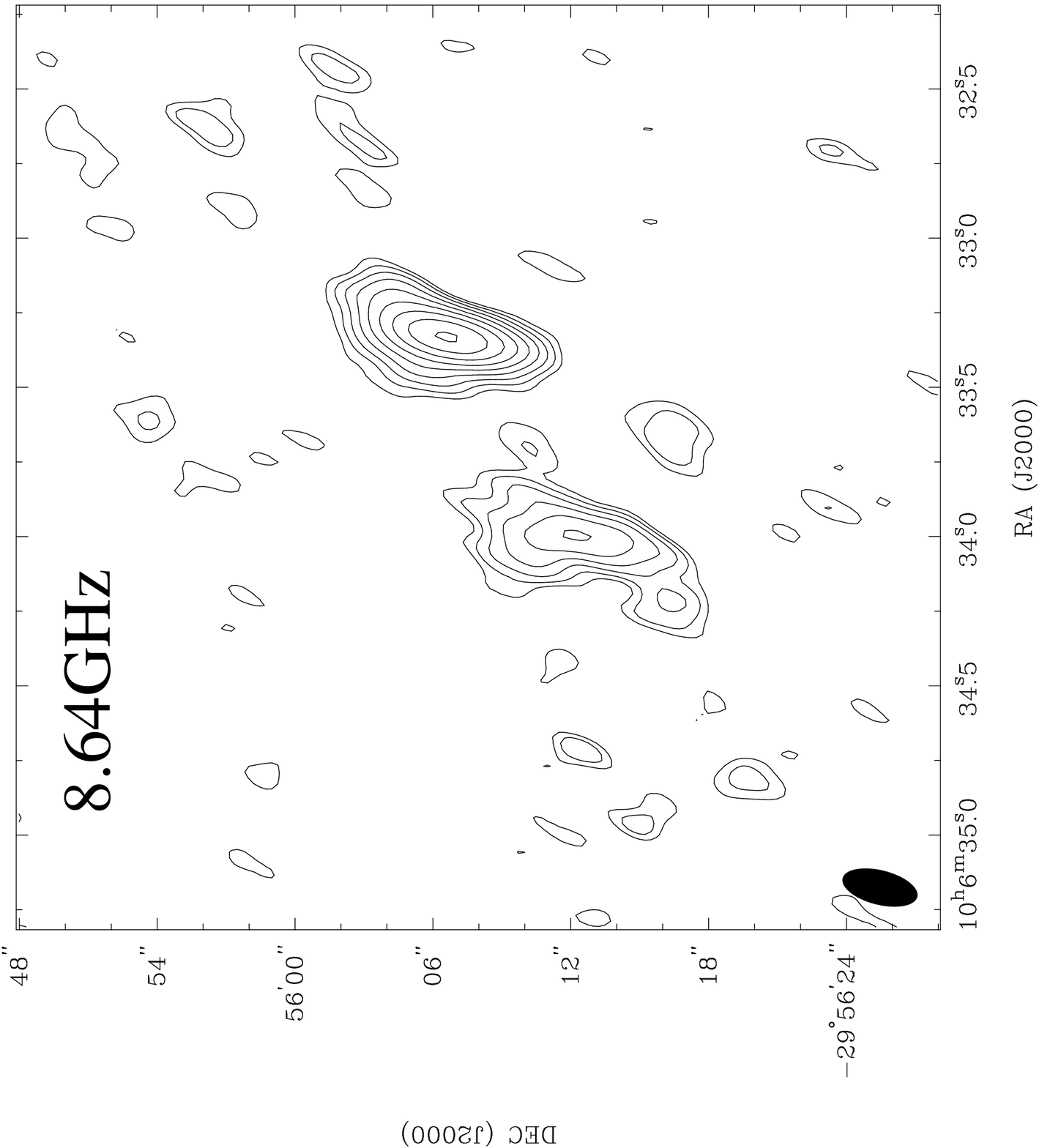}
\caption{Left: The radio morphology of NGC\,3125 at 4.80GHz. The beam
size for the 4.80GHz image is $6.06''\times 2.66''$, with a beam
position angle of $-14.3^\circ$ (shown in bottom left of image). The
$\sigma_{RMS}$ for this image is $58\mu$Jy/beam and the contours are at
$2^{n/2}\times \sigma_{RMS}$ for $n=2.....11$.  Right: The radio
morphology of NGC\,3125 at 8.64GHz.  The beam size for the 8.64GHz is
$3.34''\times 1.50''$ with a beam position angle of $-14.0^\circ$ (shown
in bottom left of image). The $\sigma_{RMS}$ for this image is
$60\mu$Jy/beam and the contours are at $2^{n/2}\times \sigma_{RMS}$ for
$n=2.....10$.}
\label{fig2}
\end{figure*}

The radio images of NGC\,3125 at 4.80GHz and 8.64GHz are shown in
Fig.~\ref{fig2}. The beam sizes for these observations are $6.06''\times
2.66''$ at 4.80GHz and $3.34''\times 1.50''$ at 8.64GHz. For the 8.64GHz
observations the beam-size corresponds to a size of $250\times 100$ pc
at the assumed distance of 13.8Mpc.  Note, that because of the antenna
failure the spatial resolution of the NGC\,3125 observations are lower
than those presented for NGC\,5408.

The observed radio morphology shown in Fig.~\ref{fig2}) consists of two
main emission regions or knots, which we designate NGC\,3125:A and
NGC\,3125:B. Taking the position from the 8.64GHz data as the most
reliable, the sources are located at (RA, Dec)$=10^h06^m 33.3^s,
-29^\circ56' 06.6''$ (source A) and $10^h06^m 33.99^s, -29^\circ 56'
12.25''$ (Source B), and are separated by $\sim 8''$ .

Using the {\small MIRIAD} {\sl imfit} package we have measured the flux
for these two sources using a Gaussian fit for each, and the results are
shown in Table~\ref{tab3}. The fitting suggests that the sources are at
most only marginally larger than the beam size.  The derived flux
densities at 4.80GHz for the two sources are 5.9mJy (NGC\,3125:A) and
4.3mJy (NGC\,3125:B), and the spectral indices are $\alpha=-0.5$, for
NGC\,3125:A and $\alpha=-0.6$ for NGC\,3125:B.

We can compare the positions of these radio sources with the optical
star clusters in NGC\,3125. As noted by Marlowe \etal\ (1995) there are
three peaks in the H$\alpha$ surface brightness within the
central 1.7kpc corresponding to star clusters. We can obtain more
accurate positions of these star clusters using archival 
{\sl HST} Faint Object Camera, and if we designate these clusters
NGC\,3125:SSC1--3, the locations of the optical SSCs are (RA, Dec
(J2000)):

\noindent NGC\,3125:SSC1: $10^h06^m33.29^s, -29^\circ 56'07.8''$\\
\noindent NGC\,3125:SSC2: $10^h 06^m 33.90^s, -29^\circ 56'12.8''$ \\
\noindent NGC\,3125:SSC3: $10^h 06^m 33.92^s, -29^\circ 56'12.9''$

\noindent This means that radio source NGC\,3125:A is likely associated
with SSC1 and radio source NGC\,3125:B is associated with either or possibly
both SSC2 and SSC3 (these two objects being separated by only
$0.2''$). It is worth noting that in the {\sl FOC} image other fainter
star-clusters are visible in NGC\,3125.

\subsection{NGC\,5408}
\label{sec4p2}

\begin{figure*}
\vspace{9cm}
\includegraphics{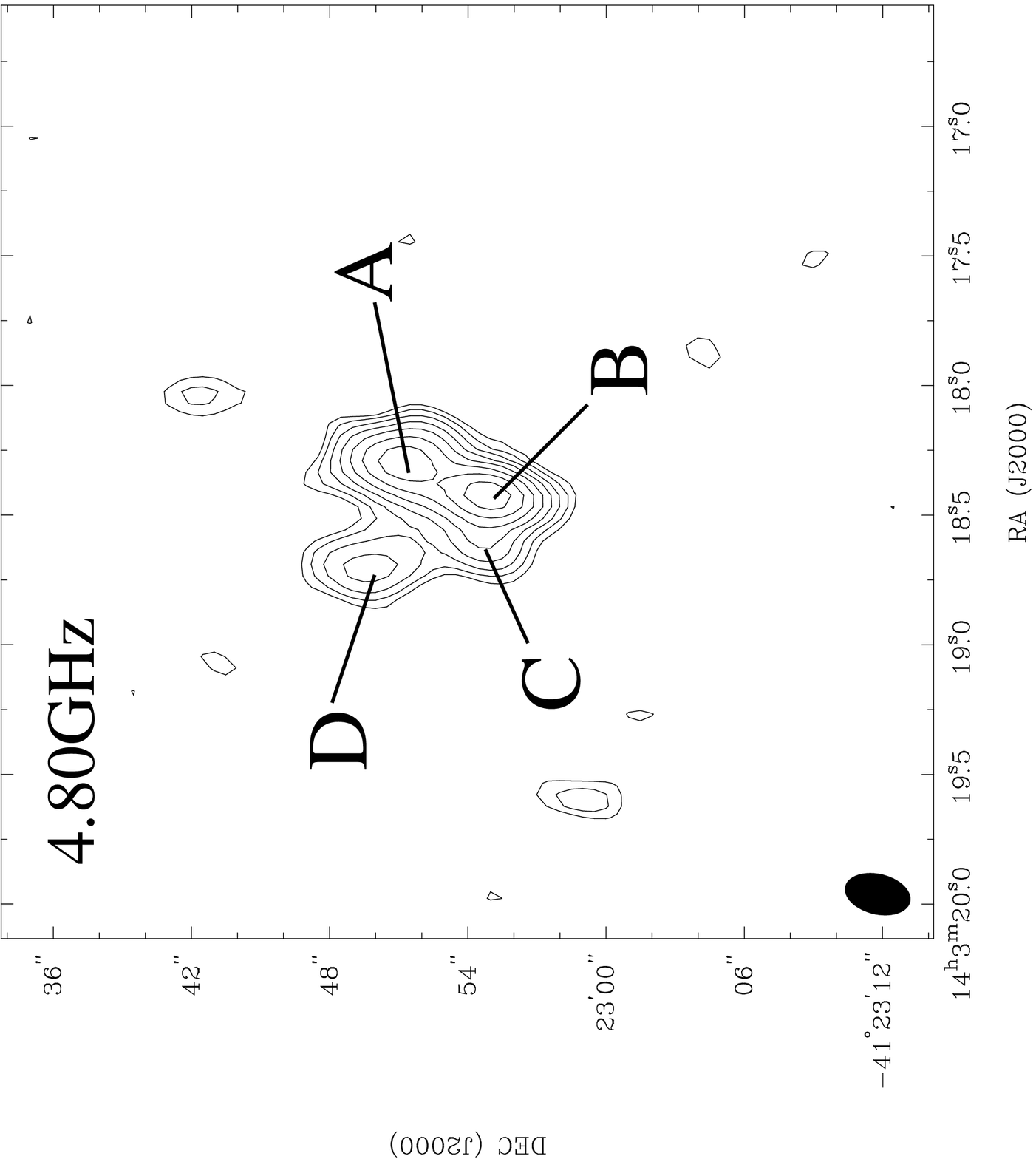}
\includegraphics{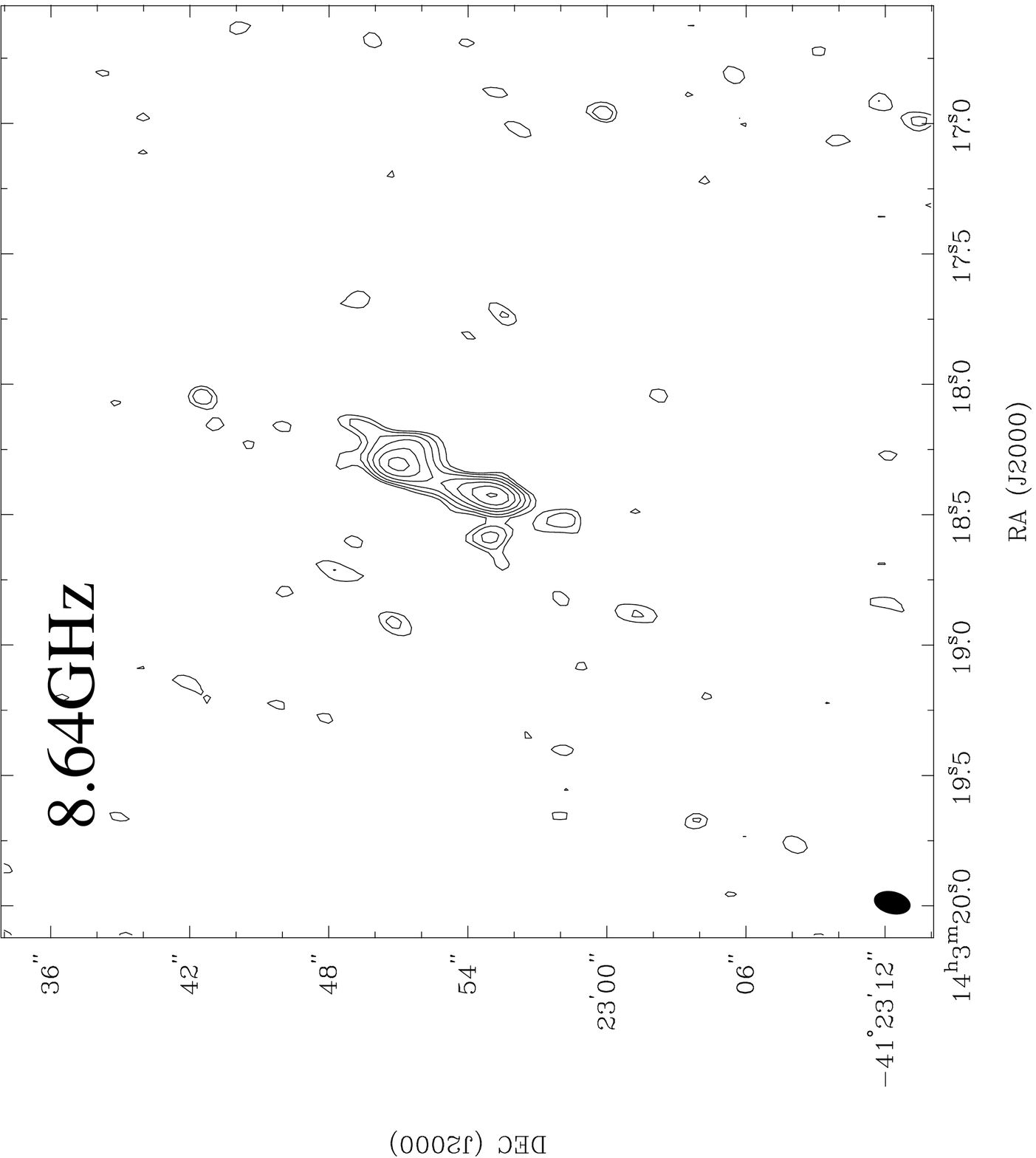}
\caption{Left: The radio morphology of NGC\,5408 at 4.80GHz The beam
size for the 4.80GHz image is $2.90''\times 1.76''$, with a beam
position angle of $-13.5^\circ$ (shown in bottom left of image). The
$\sigma_{RMS}$ for this image is $48\mu$Jy/beam and the contours are at
and the contours are at $2^{n/2}\times \sigma_{RMS}$ for $n=3.....10$.
Right: The radio morphology of NGC\,5408 at 8.64GHz.  The beam size for
the 8.64GHz is $1.60''\times 0.97''$ with a beam position angle of
$-13.3^\circ$ (shown in bottom left of image). The $\sigma_{RMS}$ for
this image is $48\mu$Jy/beam and the and the contours are at
$2^{n/2}\times \sigma_{RMS}$ for $n=3.....9$.}
\label{fig3}
\end{figure*}

The radio continuum images of NGC\,5408 at 4.80GHz and 8.64GHz are shown
in Fig.~\ref{fig3}. The beam size for these observations are
$2.90''\times 1.76''$ at 4.80GHz and $1.60''\times 0.97''$ at 8.64GHz.
It is clear that NGC\,5408 has a somewhat more complex radio morphology
than NGC\,3125, with four discrete sources, which we shall designate
NGC\,5408:(A--D), in order of increasing RA. All four sources are
apparent in the 4.80GHz image, with NGC\,5408:C showing up as an
extension to the brightest source (NGC\,5408:B).  NGC\,5408:C is more
apparent in the higher resolution 8.64GHz image, while NGC\,5408:D is
only clearly seen in the 4.80GHz map.

Again, using the {\small MIRIAD} {\sl imfit} package we have measured
the flux densities of these sources, and these and the spectral indices
for the 4 discrete sources and these are listed in
Table~\ref{tab3}. With the exception of source A, all of the spectral
indices are strongly negative. For source A the index is much flatter.

We can compare the positions of the radio sources with those of optical
star-clusters. Archival {\sl HST} {\sl WFPC2} data reveals several
compact regions of star-formation within NGC\,5408, with positions (RA,
Dec. (J2000)):

\noindent NGC\,5408:SSC1: $14^h 03^m 18.34^s, -41^\circ 22' 51.5''$\\
\noindent NGC\,5408:SSC2: $14^h 03^m 18.44^s, -41^\circ 22' 53.5''$\\ 
\noindent NGC\,5408:SSC3: $14^h 03^m 18.73^s, -41^\circ 22' 49.6''$

\noindent The regions SSC1 and SSC2 may be composed of more than one
star-cluster. The radio source A is associated with SSC1, sources B and
C appear to be unrelated to the star-forming regions, and source D is
associated with SSC3. Again, there are a number of fainter objects in
the {\sl WFPC2} image, which could be lower mass clusters.

\begin{table*}
\caption{The radio sources in NGC\,3125 and NGC\,5408. The observed
positions and the fluxes at 4.80GHz and 8.64GHz of each emission region
are shown, along with the spectral index. The numbers shown in
parentheses are the estimated errors on the source fluxes and the
spectral index.}
\begin{tabular}{lccccc}\hline
Source & RA &  Dec & $S_{4.80}$ & $S_{8.64}$ & Spectral Index \\ 
       & \multicolumn{2}{c}{(J2000)} & (mJy) & (mJy) &($\alpha$)\\ \hline
NGC\,3125:A & 10 06 33.33 & $-$29 56 06.6 & $5.9 \ (0.06)$ & $4.5\ (0.06)$ &
$-0.5\ (0.03)$\\
NGC\,3125:B & 10 06 33.99 & $-$29 56 12.2 & $4.3 \ (0.06)$ & $3.0\ (0.06)$ &
$-0.6\ (0.04)$\\
\hline
NGC\,5408:A & 14 03 18.30 & $-$41 22 51.2 & $1.9\ (0.05)$ & $1.8\ (0.05)$ &
$-0.1\ (0.07)$\\
NGC\,5408:B & 14 03 18.43 & $-$41 22 54.9 & $2.2\ (0.05)$ & $1.6\ (0.05)$ &
$-0.6\ (0.07)$\\
NGC\,5408:C & 14 03 18.59 & $-$41 22 54.9 & $0.7\ (0.05)$ & $0.5\ (0.05)$ &
$-0.5\ (0.21) $\\
NGC\,5408:D & 14 03 18.69 & $-$41 22 49.8 & $0.6\ (0.05)$ & $0.4\ (0.05)$ &
$-0.8\ (0.26)$\\
\hline
\end{tabular}
\label{tab3}
\end{table*}

\subsection{Discussion}
\label{sec4p3}

The observed radio morphologies of these two dwarf starbursts shows
several discrete regions of emission, in which much of the observed
emission is coincident with massive star-clusters in both galaxies.
Leaving aside the possibility of an AGN and emission mechanisms
associated with a massive black-hole (for which there appears to be no
evidence of in these galaxies) the likely emission mechanisms are
(c.f. Forbes \& Norris 1998):

\begin{enumerate} 

\item Supernovae and supernova remnants: cosmic rays accelerated by
supernovae and supernova remnants, which then interact with the
interstellar magnetic field to emit synchrotron emission. The typical
spectral index of this synchrotron emission will be $\alpha \sim
-0.7$. In starburst and normal galaxies this component will tend to be
distributed throughout the disk.  However, in NGC\,3125 and NGC\,5408 we
are dealing with dense stellar clusters, which could contain many SN
remnants in a small volume, giving rise to a compact region of
synchrotron emission.

\item \hii\ regions: hot electrons in \hii\ regions generate free--free
emission. At cm wavelengths most \hii\ regions are optically thin,
resulting in a flat spectrum ($\alpha=-0.1$). However, some compact
H{\small II} regions are seen to be optically thick, giving rise to a
rising spectrum at cm wavelengths (NGC\,5253 -- Turner, Ho \& Beck 1998;
Hen~2-10 -- Kobulnicky \& Johnson 1999; NGC\,2146 -- Tarchi \etal\ 200;
M82 -- McDonald \etal\ 2002). In these cases, which are probably
associated with very young super star clusters, the gas densities are
sufficient to cause a turn-over in the emission at low frequencies.
In all \hii\ regions the radio spectrum does turnover at some
sufficiently low frequency (for instance, for Orion this occurs at
$\lambda=1$m).

\item The radio emission from ultra--luminous infrared galaxies is
optically thick to free--free absorption, so that the typical
synchrotron spectrum of these galaxies is flattened at low frequencies
(Condon \etal\ 1991).

\end{enumerate}

The combined result of these effects in Seyfert and starburst galaxies
is to produce a typical radio spectral index of $\alpha\sim -0.7$ (from
the extended synchrotron emission) with a flattening at low frequencies
in some starburst sources because of free--free absorption.

For NGC\,3125 and NGC\,5408, the morphology and spectral indices of the
radio knots imply that the observed emission is primarily associated
with type II supernovae activity which is associated with star-formation
activity within the star-clusters. The exception is NGC\,5408:A, which
has a flatter index indicative of optically thin free-free emission. The
obvious interpretation is that in this source, which is associated with
an SSC, the emission is dominated by the $\hii$ region, and that the
cluster is perhaps younger than the others in NGC\,5408 and supernova
activity has not yet begun to dominate the radio emission.

The next step is to see if we can develop a coherent picture of the
star-forming behaviour of these galaxies. It is worth noting that in
these galaxies the optical light may be dominated (or have a substantial
contribution) from the star-clusters, but it is likely that the stellar
content these clusters only comprise a few percent of the total stellar
mass in the galaxies. However, these clusters may provide a substantial
fraction of the ongoing star-formation in the galaxy.

We can estimate the expected supernova rate in the radio emitting
regions. Assuming that the 4.80GHz flux is non-thermal emission, then
the implied SN rate in the various radio emitting regions can be estimated
from (Condon \& Yin 1990):

\begin{equation}
L_{\rm 6cm} ({\rm W~Hz}^{-1})= 1.3\times 10^{23}  \left(\frac{\nu}{\rm
1GHz}\right)^{-0.8} \nu_{SN} (\peryr)
\end{equation}

For NGC\,3125 this corresponds to SN rate of $\nu_{SN}=4\times 10^{-3}\peryr$
for NGC\,3125:A and $\nu_{SN}=3\times 10^{-3}\peryr$ for NGC\,3125:B. The
corresponding SN rates for the NGC\,5408 regions are in the range
$\nu_{SN}=(1-5)\times 10^{-4}\peryr$.

Similarly, we can estimate the star-formation rates from both
far-infrared (FIR) observations and these radio observations. The FIR
rates will apply to the whole galaxy while the radio ones will apply to
the observed emission regions (see Hopkins \etal\ 2001 for a comparison
of the various methods for estimating star-formation rates in galaxies)

The FIR luminosities of $L_{FIR}=4.0\times 10^{42}\ergs$ (NGC\,3125) and
$L_{FIR}=7.4\times 10^{41}\ergs$ (NGC\,5408). From Kennicutt (1998) we
can estimate the total star-formation rate (SFR) in each galaxy from the
following relationship:

\begin{equation} 
{\rm SFR} (\msunyr) = 4.5\times 10^{-44} L_{FIR}
(\ergs) 
\end{equation} 

\noindent where the mass range for the star-formation rate is assumed to
be $0.1-100M_\odot$ and $L_{FIR}$ is the far infrared luminosity. This
means that the estimated total SFR in the two galaxies are $0.18
\msunyr$ (NGC\,3125) and $0.033\msunyr$ (NGC\,5408). Note, that assuming
a Salpeter IMF, the star-formation rate of high mass, that is stars with
$M_i>10 M_\odot$, will be about 10 per cent of this value.

Using the results of Condon (1992, see also Haarsma \etal\ 2000) we can
estimate the star-formation rates in the radio regions from these radio
observations. The implied total star-formation rate in the two regions
in NGC\,3125 is $0.3\msunyr$ and in NGC\,5408 the figure is
$0.09\msunyr$, somewhat higher than the FIR values.

As an aside, we can compare the estimated SN rates with the theoretical
expectations from Leitherer \& Heckman (1995). For an ongoing starburst
with a star-formation rate of $1\msunyr$, a Salpeter IMF and solar
abundance, the expected supernova rate is $\nu_{SN}\sim 0.02\peryr$. For
the SFR rates estimated above, this corresponds to supernova rates of
$\nu_{SN}=3.6\times 10^{-3}\peryr$ for NGC\,3125 and $6.6\times
10^{-4}\peryr$ for NGC\,5408. Comparing these values to the estimated SN
rates above, there is reasonable agreement between the integrated SN
rates in the various radio emitting regions and these expectations.

As discussed earlier, both galaxies are also X-ray luminous. There are
several potential mechanisms for generating the X-rays; an AGN, X-ray
binaries, SNRs and emission from superbubbles. There is no evidence for
an AGN, and while individual supernovae or SNRs can be very radio and
X-ray luminous (van Dyk \etal\ 1993; Fabian \& Terlevich 1996; Immler \&
Lewin 2002), their duration as high luminosity object is
short-lived. Consequently, high mass X-ray binaries, associated with the
active star-formation in these galaxies, are likely to provide much of
the observed X-ray emission.

Extrapolating from the star-formation rate in our own Galaxy ($\sim
0.1-0.2\msunyr$) we can estimate the number of luminous X-ray binaries
in each system. The Galaxy has an estimated population of luminous high
mass X-ray binaries (HMXBs) of $\sim 27\pm 19$ (Dalton \& Sarazin 1995).
Based on this, these two dwarf starbursts should have a high mass X-ray
binary population of $20\pm 10$ for NGC\,3125 and $5\pm 2$ for
NGC\,5408.  Note, that this refers only to the most luminous HMXBs,
namely those with $L_X\geq 10^{37}\ergs$.

The observed X-ray luminosities of these galaxies are $L_X=2\times
10^{39}\ergs$ (NGC\,3125) and $3\times 10^{40}\ergs$ (NGC\,5408),
meaning that NGC\,5408 is substantially more X-ray luminous than
NGC\,3125, although having a smaller star-formation rate. The X-ray
emission from NGC\,3125 could be dominated by X-ray binaries, however,
this is unlikely to be the case for NGC\,5408 and we may need an
additional very X-ray luminous source, such as the objects seen by the
\chandra\ satellite in other star-forming galaxies (for instance, Kaaret
\etal\ 2001) believed to be intermediate mass black-holes.

Consequently, from these observations, we can conclude that the radio
emission in NGC\,3125 and NGC\,5408 is predominantly non-thermal in
nature and is most likely dominated by processes associated with type II
SNRs which are associated with recent starburst activity in the massive
stellar clusters. Only one source in NGC\,5408 has a flatter spectral
index indicative of optically thin free-free emission. It will remain to
be seen if the cluster associated with this radio source is younger than
the other clusters in this galaxy.  In the case of NGC\,3125 we have a
coherent picture, whereby the expected star-formation rate, supernovae
rate and massive X-ray binaries populations can give rise to the
observed properties, while in the case of NGC\,5408 although the FIR and
radio luminosities are broadly consistent the galaxy is substantially
X-ray overluminous. This galaxy may well be the home of a single
super-luminous system such as has been seen in several other galaxies.

One problem with the simple analysis presented here is that it has mostly
assumed a constant star-formation rate. In many of these stellar clusters
there may be a coeval population of stars and in this situation the SN
rate can be very time variable. A more detailed optical analysis is
needed for these galaxies, to investigate the masses/ages of the
star-clusters. However, these new observations are throwing some more
light on the radio evolution of massive star-clusters in dwarf starbursts.

In conclusion, we have presented new radio observations of two dwarf
starburst galaxies, NGC\,3125 and NGC\,5408. Several discrete regions of
radio emission are seen in these galaxies, most of them associated with
massive young star clusters. The radio spectral indices indicate the
emission is dominated by synchrotron emission associated with supernova
activity, though in one case a flatter spectrum is observed, possibly
indicative of a younger cluster.

\section*{Acknowledgements}

The Australia Telescope is funded by the Commonwealth of Australia for
operation as a National Facility funded by CSIRO. Ian Stevens is
supported by a PPARC Advanced Fellowship.


\begin{thebibliography}{99}

\bibitem{}Beck S.C., Turner J.L., Kovo O., 2000, AJ, 120, 244

\bibitem{}Bohuski T.J., Burbidge E.M., Burbidge G.R., Smith M.G., 1972,
ApJ, 175, 329

\bibitem{}Condon J.J., Yin Q.F., 1990, ApJ, 357, 97

\bibitem{}Condon J.J., Huang Z.P., Yin Q.F., Thuan T.X., 1991, ApJ, 378, 65

\bibitem{}Condon J.J., 1992, ARA\&A, 30,575

\bibitem{} Conti P.S., 1991, ApJ, 377, 115

\bibitem{dalton} Dalton W.W., Sarazin C.L., 1995, ApJ, 440, 280

\bibitem{}Dekel A., Silk J., 1986, ApJ, 303, 39 

\bibitem{}Fabian A.C., Ward M.J., 1993, MNRAS, 263, L51

\bibitem{}Fabian A.C., Terlevich R.J., 1996, MNRAS, 280, L5

\bibitem{}Forbes D.A., Norris R.P., 1998, MNRAS, 300, 757

\bibitem{}Fourniol N., Pakull M., Motch C., 1996, In: \lq\lq
Rontgenstrahlung from the Universe\rq\rq\, eds. Zimmermann H.U., Trumper
J., and Yorke H., MPE Report 263, p. 375

\bibitem{}Greve A., Tarchi A., Huttmeister S., de Grijs R., van der
Hulst J.M., Garrington S.T., Neininger N., 2002, A\&A, 381, 825

\bibitem{}Haarsma D.B., Partridge R.B., Windhorst R.A., Richards E.A.,
2000, ApJ, 544, 641


\bibitem{}Hopkins A.M., Connolly A.J., Haarsma D.B., Cram L.E., 2001,
AJ, 122, 288

\bibitem{}Immler S., Lewin W.H.G., 2002, astro-ph/0202231.

\bibitem{}Johnson K.E., Kobulnicky H.A., Massey P., Conti P.S., 2001,
ApJ, 559, 864


\bibitem{} Kaaret P., Prestwich A.H., Zezas A., Murray S.S., Kim D.-W.,
Kilgard R.E., Schlegel E.M., Ward M.J., 2001, MNRAS, 321, L29

\bibitem{kenn} Kennicutt R.C., 1998, ARA\&A, 36, 189

\bibitem{}Klein U., Wielebinski R., Thuan T.X., 1984, A\&A, 141, 241

\bibitem{}Kobulnicky H.A., Johnson K.E., 1999, ApJ, 527, 154

\bibitem{}Kobulnicky H.A., Johnson K.E., 2000, ApJ, 539, 1023

\bibitem{}Kobulnicky H.A., Johnson K.E., 2001, In: 
\lq\lq Starbursts: Near and Far,\rq\rq (Springer-Verlag), p.95

\bibitem{}Kronberg P.P., Sramek R.A., Birk G.T., Dufton Q.W., Clarke
T.E., Allen M.L., 2000, ApJ, 535, 706

\bibitem{}Kunth  D., Maurogordato S., Vigroux L., 1988, A\&A, 204, 10

\bibitem{}Leitherer C., Heckman T.M., 1995, ApJS, 96, 9

\bibitem{}Mac Low M.-Mark.,  Ferrara A., 1999, ApJ, 513, 142

\bibitem{}Marlowe A.T., Heckman, T.M., Wyse, R.G., Schommer R., 1995,
ApJ, 438, 563

\bibitem{}Marlowe A.T., Meurer G.R., Heckman, T.M., 1999,
ApJ, 522, 183

\bibitem{}McDonald A.R., Wills K.A., Muxlow T.W.B., Pedlar A., 2002,
MNRAS, submitted

\bibitem{}Raimann D., Bica E., Storchi-Bergmann T., Melnick J., Schmitt
H., 2000, MNRAS, 314, 295

\bibitem{}Sault R.J., Teuben P.J., Wright M.C.H., 1995, in Astronomical
Data Analysis Software and Systems IV, ed. R. Shaw, H.E. Payne,
J.J.E. Hayes, ASP Conf. Ser., 77, p.433.

\bibitem{}Stevens I.R., Strickland D.K., 1998a, MNRAS, 294, 523

\bibitem{}Stevens I.R., Strickland D.K., 1998b, MNRAS, 301, 215

\bibitem{}Strickland D.K., Stevens I.R., 2000, MNRAS, 314, 511

\bibitem{}Tarchi A., Neininger N., Greve A., Klein U., Garrington S.T.,
Muxlow T.W.B., Pedlar A., Glendenning B.E., 2000, A\&A, 358, 95

\bibitem{}Turner J.L., Ho P.T.P., Beck S.C., 1998, AJ, 116, 1212

\bibitem{}Turner J.L., Beck S.C., Ho P.T.P., 2000, ApJ, 532, L109

\bibitem{}Vacca W.D., Conti P.S., 1992, ApJ, 401, 543

\bibitem{}Van Dyk S.D., Weiler K.W., Sramek R.A., Panagia N., 1993, 
ApJ, 419, L69

\bibitem{}Venturi T., Bardelli S., Zambelli G., Morganti R., Hunstead
R.W., 2001, MNRAS, 324, 1131

\bibitem{}Whitmore B.C., 2002, In: STScI Symposium Series 14
(ed. M. Livio), in press

\end{thebibliography}
\end{document}